\documentclass[twocolumn,superscriptaddress,floatfix]{revtex4}

\begin{document}

\title{Non-Linear Stochastic Equations with Calculable Steady States}
\author{Rava da Silveira}
\affiliation{Lyman Laboratory of Physics, Harvard University, Cambridge, Massachusetts 02138, U.S.A.\\
and Laboratoire de Physique Th\'eorique, Ecole Normale Sup\'erieure, 24 rue Lhomond, 
75005 Paris, France}
\author{Mehran Kardar}
\affiliation{Department of Physics, Massachusetts Institute of Technology, Cambridge, Massachusetts 02139, U.S.A.}

\begin{abstract}
We consider generalizations of the Kardar--Parisi--Zhang equation that
accomodate spatial anisotropies and the coupled evolution of several fields,
and focus on their symmetries and non-perturbative properties. In
particular, we derive generalized fluctuation--dissipation conditions on the
form of the (non-linear) equations for the realization of a Gaussian
probability density of the fields in the steady state. For the amorphous
growth of a single height field in one dimension we give a general class of
equations with exactly calculable (Gaussian and more complicated) steady
states. In two dimensions, we show that any anisotropic system evolves on
long time and length scales either to the usual isotropic strong coupling
regime or to a linear-like fixed point associated with a hidden symmetry.
Similar results are derived for textural growth equations that couple the
height field with additional order parameters which fluctuate on the growing
surface. In this context, we propose phenomenological equations for the
growth of a crystalline material, where the height field interacts with
lattice distortions, and identify two special cases that obtain Gaussian
steady states. In the first case compression modes influence growth and are
advected by height fluctuations, while in the second case it is the density
of dislocations that couples with the height.
\end{abstract}

\maketitle

\section{Introduction}

Non-linear stochastic partial differential equations appear extensively in
problems of equilibrium and non-equilibrium statistical physics. For systems
in thermal equilibrium, the form of these equations is constrained by
fluctuation--dissipation conditions \cite{Deker} that ensure convergence of
the steady state probability distribution to the appropriate Boltzmann
weight \cite{HH}. Non-equilibrium systems are not similarly constrained, and
there is no simple way of finding their behavior in steady state (if any).
However, there are examples in which steady states can be found exactly as
solutions of the associated Fokker--Planck equations. In this paper we
review some such examples, and introduce several new ones, along the way
seeking general principles for finding steady states associated to
non-linear stochastic equations exactly.

The simplest equation, which serves as the prototype for our investigations,
is the Kardar--Parisi--Zhang (KPZ) equation \cite{KPZ} 
\begin{equation}
\partial _{t}h=\nu \nabla ^{2}h+\frac{1}{2}\lambda \left( \nabla h\right)
^{2}+\eta ,  \label{KPZ}
\end{equation}%
describing the non-equilibrium fluctuations of a (height) field $h(\mathbf{x}%
,t)$. The equation is equivalent to the Burgers equation (for a field $%
\mathbf{u}=\nabla h$) for vorticity-free turbulence \cite{Forster}, and
appears in various guises in the study of domain walls \cite{HHF} and
directed polymers \cite{KZ,FH} in a random potential, surface growth \cite%
{Reviews}, and even the gene (or protein) sequence alignment problem \cite%
{HL,Hwa}. Still more problems can be formulated as generalizations of the
KPZ equation that accommodate spatial anisotropy or the interplay of several
fields. Examples in the literature include the dynamics of a vicinal surface 
\cite{Villain,Wolf}, the growth of two coupled surfaces \cite{Barabasi} or
of a magnetic surface \cite{Drossel,DK,Kardar99}, and the transport of a
flux line or polymer \cite{Ertas}.

The stochastic aspect of Eq.~(\ref{KPZ}) is due to the noise $\eta (\mathbf{x%
},t)$, which has zero mean and short-range correlations in space and time.
In the absence of the non-linear term proportional to $\lambda $, it reduces
to a standard Langevin equation, with a Gaussian steady state. In one
dimension, the non-linear term does not modify this steady state as the
associated probability current in the Fokker--Planck equation is zero. This
observation motivates our search for other equations with this property,
namely an easily guessed (equilibrium) steady state which is not affected by
the additional (non-equilibrium) non-linearities.

In Sec.~\ref{1d}, we start by constructing the Fokker--Planck equation for
the one dimensional KPZ equation and explicitly showing that the probability
current due to the non-linear term does not modify the steady state, as it
appears in an integral of a complete derivative. This observation is then
used as a basis for constructing other one dimensional non-linear equations
that share this property. Indeed, we find that the class of such equations
is quite large, including some equations already considered in the
literature.

Higher dimensional versions of the KPZ equation may also obtain a Gaussian
steady state in spite of their non-linear character. We discuss such a case
in Sec.~\ref{2d}, namely an anisotropic variant of the KPZ equation in two
dimensions with non-linear terms of opposite signs in the two directions.
Using renormalization group methods, Wolf \cite{Wolf} showed that this model
indeed flows under renormalization to a linear fixed point. Generalizations
of this equation with calculable steady states are also constructed; they
all share a hidden symmetry under reflection, absent in the isotropic KPZ
equation.

The examples from one and two dimensions motivate the search for more
general principles governing the existence of simple steady states, taken up
in Sec.~\ref{general}. Specifically, we consider stochastic dynamics of
multiple fields coupled by non-linear (possibly anisotropic) generalizations
of the KPZ term, and ask whether they admit Gaussian steady states. A direct
solution of the Fokker--Planck equation becomes considerably more difficult
and, instead, we derive two sets of general prescriptions on the
coefficients for this to occur. These prescriptions may be viewed as
generalized fluctuation--dissipation relations \cite{glasses}, and are quite
restrictive. In particular, they cannot be satisfied in three and higher
dimensions as we show in Sec. \ref{3d}.

Having obtained general prescriptions, in Sec.~\ref{coupled} we apply them
to equations for coupled fields in one and two dimensions. Some of the
examples we discuss correspond to equations that have already appeared in
the literature, in particular pertaining to the dynamics of a flux--line or
polymer (Sec.~\ref{lines}) and to the growth of a magnetic film (Sec.~\ref%
{magnetic}). However, in Sec.~\ref{Xtal} we propose a novel set of equations
to describe the coupling of the strain field of a growing crystal to its
height fluctuations. We find that Gaussian steady states are indeed
permitted for these equations in special cases.

The Appendix treats a simple example aimed at illustrating how the
systematic approach may be extended to exactly calculable non-Gaussian
steady states.

\section{Generalized Growth Equations in One Dimensions}

\label{1d} Consider the probability distribution $\mathcal{P}[h]$ for
configurations of the field $h(\mathbf{x})$. As the surface changes in time
according to Eq.~(\ref{KPZ}), the corresponding probability evolves
according to the Fokker--Planck equation 
\begin{equation}
\partial _{t}\mathcal{P}=\int d^{d}x\left\{ -\left[ \nu \nabla ^{2}h+\frac{1%
}{2}\lambda \left( \nabla h\right) ^{2}\right] \frac{\delta \mathcal{P}}{%
\delta h}+D\frac{\delta ^{2}\mathcal{P}}{\delta h^{2}}\right\} .
\label{FP-KPZ}
\end{equation}%
The term in the square brackets is due to the deterministic probability
current and the remainder comes from the stochastic noise, assumed Gaussian
with $\langle \eta (\mathbf{x},t)\rangle =0$ and 
\begin{equation}
\left\langle \eta (\mathbf{x},t)\eta (\mathbf{x}^{\prime },t^{\prime
})\right\rangle =2D\delta ^{d}(\mathbf{x}-\mathbf{x}^{\prime })\delta
(t-t^{\prime }).  \label{noise}
\end{equation}%
In equilibrium, $D=k_{B}T$; more generally, $D\,$\ is a measure of the
magnitude of the noise.

A steady-state solution is one for which $\partial _{t}\mathcal{P}=0$. In
the absence of the non-linear term, the steady-state solution of Eq.~(\ref%
{FP-KPZ}) is a simple Gaussian, 
\begin{equation}
\mathcal{P}=N\exp \left[ -\frac{\nu }{2D}\int d^{d}\mathbf{x}(\nabla h)^{2}%
\right] ,  \label{Gaussian}
\end{equation}%
where $N$ is a normalization constant. In general, this is not a steady
state for $\lambda \neq 0$. In one dimension, however, the contribution of
the non-linear term to the probability current can be simplified to 
\begin{eqnarray}
\partial _{t}\mathcal{P} &=&-\mathcal{P}\int dx\frac{\nu \lambda }{2D}%
(\partial _{x}h)^{2}\partial _{xx}h  \nonumber \\
&=&-\mathcal{P}\int dx\partial _{x}\left[ \frac{\nu \lambda }{6D}(\partial
_{x}h)^{3}\right] ,
\end{eqnarray}%
a surface integral safely set to zero in the limit of an infinite system.
Thus, the steady-state spatial correlations are not influenced by the
presence of the KPZ non-linearity and coincide with the Gaussian
(Edwards--Wilkinson \cite{EW}) ones. We make no statements about the
stability of the steady state of Eq. (\ref{Gaussian}); however, here, and in
every other example for which simulations are available, numerical results
indicate that the simple Gaussian steady states we discuss are indeed the
ones achieved at long time.

The one-dimensional KPZ equation is a particular instance of a more general
class of equations, 
\begin{equation}
\partial _{t}h=f\left( {\partial _{x}}h\right) {\partial _{x}}{\partial _{x}}%
h+g\left( {\partial _{x}}h\right) +\eta ,  \label{1d-mgen}
\end{equation}%
with $f$ and $g$ arbitrary functions. These equations obtain a steady state 
\begin{equation}
\mathcal{P}=N\exp \left[ -\frac{1}{D}\int dxF({\partial _{x}}h)\right] ,
\end{equation}%
where $f$ is the second derivative of $F$ (\emph{i.e.}, ${d}^{2}{F(u)}/{du}%
^{2}=f(u)$). For $g(\partial _{x}h)=0$, Eq.~(\ref{1d-mgen}) is again a
standard Langevin equation, while the contribution of this function to the
probability current is 
\begin{eqnarray}
\partial _{t}\mathcal{P} &=&-\mathcal{P}\int dx\frac{1}{D}g({\partial _{x}}%
h)f({\partial _{x}}h){\partial _{x}}{\partial _{x}}h  \nonumber \\
&=&-\mathcal{P}\int dx\partial _{x}\left[ \frac{\nu }{D}G({\partial _{x}}h)%
\right]   \nonumber \\
&=&0,
\end{eqnarray}%
where G is the primitive of $gf$ (\emph{i.e.}, ${dG(u)}/{du}=g(u)f(u)$).

The special case of a cubic non-linearity $g({\partial _{x}}h)=\frac{1}{2}%
\lambda (\partial _{x}h)^{2}+\frac{1}{6}\lambda ^{\prime }(\partial
_{x}h)^{3}$ (with $f=1$) was introduced to describe an interface separating
stationary phases of the Toom model \cite{DLSS}. The Gaussian steady state
corroborates the marginal irrelevance \cite{PBMH} of the cubic term, and
implies spatial correlations of the form 
\begin{equation}
\langle (h(x)-h(x^{\prime }))^{2}\rangle ^{1/2}\sim |x-x^{\prime }|^{1/2}.
\end{equation}

\section{Anisotropic Equations in Two Dimensions}

\label{2d}

To describe the growth of a vicinal (slightly miscut from a low index facet)
surface, Villain introduced \cite{Villain} an anisotropic version of the KPZ
equation, which was subsequently studied with a renormalization group
calculation by Wolf \cite{Wolf}. This generalized equation has the form 
\begin{equation}
\partial _{t}h=\nu _{x}\partial _{x}^{2}h+\nu _{y}\partial _{y}^{2}h+\frac{1%
}{2}\lambda _{x}(\partial _{x}h)^{2}+\frac{1}{2}\lambda _{y}(\partial
_{y}h)^{2}+\eta .  \label{vw}
\end{equation}%
Under renormalization, the subspace with $\nu _{x}/\nu _{y}=\lambda
_{x}/\lambda _{y}$ is fixed and equivalent to the isotropic KPZ equation
modulo a rescaling of $x$ or $y$. This subspace is locally attractive, so
that the equation flows to a strong-coupling limit if $\lambda _{x}$ and $%
\lambda _{y}$ have the same sign (stability requires $\nu _{x},\nu _{y}\geq
0 $). The more surprising behavior arises when the product $\lambda
_{x}\lambda _{y}$ is negative, in which case the flows converge to a fixed
point with vanishing non-linearities.

This vanishing of non-linearities at long length and time scales suggests a
Gaussian steady-state probability density, as corroborated by an exact
solution of a discrete model belonging to the same universality class \cite%
{Prahofer} and by a direct solution of the Fokker--Planck equation \cite%
{Kardar98}. Indeed, the Ansatz 
\begin{equation}
\mathcal{P}=N\exp \left\{ -\frac{1}{2D}\int dxdy\left[ \nu _{x}\left(
\partial _{x}h\right) ^{2}+\nu _{y}\left( \partial _{y}h\right) ^{2}\right]
\right\} ,
\end{equation}%
with the generalized fluctuation--dissipation condition ${\nu _{x}}/\nu
_{y}=-{\lambda _{x}}/{\lambda _{y}}$, solves for the steady state. To verify
this, we note that the contributions from the non-linearities take the form 
\begin{eqnarray}
\partial _{t}\mathcal{P} &=&-\frac{\mathcal{P}}{2D}\int dxdy\left[ \lambda
_{x}\left( \partial _{x}h\right) ^{2}+\lambda _{y}\left( \partial
_{y}h\right) ^{2}\right]  \nonumber  \label{anisoFDT} \\
&\qquad &\qquad \qquad \qquad \times \left[ \nu _{x}\partial _{x}^{2}h+\nu
_{y}\partial _{y}^{2}h\right]  \nonumber \\
&=&-\frac{\mathcal{P}}{2D}\int dxdy  \nonumber \\
&&\Big\{\partial _{x}\left[ \frac{\lambda _{x}\nu _{x}}{3}\left( \partial
_{x}h\right) ^{3}+\lambda _{y}\nu _{x}\left( \partial _{x}h\right) \left(
\partial _{y}h\right) ^{2}\right]  \nonumber \\
&&+\partial _{y}\left[ \frac{\lambda _{y}\nu _{y}}{3}\left( \partial
_{y}h\right) ^{3}+\lambda _{x}\nu _{y}\left( \partial _{x}h\right)
^{2}\left( \partial _{y}h\right) \right]  \nonumber \\
&&+2\partial _{x}h\partial _{y}h\partial _{x}\partial _{y}h\left( \lambda
_{y}\nu _{x}+\lambda _{x}\nu _{y}\right) \Big\}.
\end{eqnarray}%
If $\lambda _{y}\nu _{x}+\lambda _{x}\nu _{y}=0$, the above contribution is
the divergence of a vector field, and hence vanishes subject to the usual
boundary conditions. This non-perturbative derivation complements the
renormalization group analysis \cite{Wolf} which captures perturbatively the
character of this state at large scales and the dynamics that lead to it. In
particular, it demonstrates that a Gaussian steady state (and the
logarithmic roughness it implies) obtains at \emph{any} length scale, and
not only in the long wavelength limit.

By suitable rescalings of $x$ and $y$, we can make $\nu _{x}=\nu _{y}=\nu $,
so that the steady state reduces to Eq.~(\ref{Gaussian}) with $d=2$.
Trivially, this steady state also holds for any equation related to 
\begin{equation}
\partial _{t}h=\nu \nabla ^{2}h+\frac{\lambda }{2}\left[ \left( {\partial
_{x}}h\right) ^{2}-\left( {\partial _{y}}h\right) ^{2}\right] +\eta ,
\label{antisym}
\end{equation}%
by a rotation of the plane. This class comprises all the equations of the
form%
\begin{equation}
\partial _{t}h=\nu \nabla ^{2}h+\frac{\lambda _{1}}{2}\left[ \left( {%
\partial _{x}}h\right) ^{2}-\left( {\partial _{y}}h\right) ^{2}\right]
+\lambda _{2}{\partial _{x}}h{\partial _{y}}h+\eta ,  \label{antisym-gen}
\end{equation}%
where $\arctan (\lambda _{2}/\lambda _{1})/2$ is the plane rotation angle;
in particular, the equation 
\begin{equation}
\partial _{t}h=\nu \nabla ^{2}h+\lambda {\partial _{x}}h{\partial _{y}}h+\eta
\label{antisym-cross}
\end{equation}%
is obtained from Eq.~(\ref{antisym}) by a 45$^{\circ }$ rotation.

The surprisingly simple steady state of Eq.~(\ref{antisym}) results from a
\textquotedblleft hidden\textquotedblright\ symmetry under the transformation%
\begin{equation}
\left\{ 
\begin{array}{l}
h\rightarrow -h \\ 
x\rightarrow y \\ 
y\rightarrow x%
\end{array}%
\right. .  \label{symmetry}
\end{equation}%
The symmetry $h\rightarrow -h$ is precisely the one broken by the isotropic
KPZ non-linearity. It is restored here, provided the plane is also inverted
about an appropriate axis: the bisector in the case of Eq.~(\ref{antisym}),
the $x$- or $y$-axis in the case of Eq.~(\ref{antisym-cross}), and a
properly rotated axis in the general case of Eq.~(\ref{antisym-gen}). This
hidden symmetry sheds light on the renormalization group analysis \cite{Wolf}%
, as any two dimensional anisotropic KPZ term may be written, upon rotation
of the plane, as the sum of an isotropic part and the antisymmetric part of
Eq.~(\ref{antisym}) whose subspace is invariant under renormalization.

By analogy to Eq.~(\ref{1d-mgen}), we can generalize Eq.~(\ref{antisym-cross}%
) to include a more complicated Laplacian term, as 
\begin{equation}
\partial _{t}h=f_{x}({\partial _{x}}h){\partial _{x}}^{2}h+f_{y}({\partial
_{y}}h){\partial _{y}}^{2}h+\lambda {\partial _{x}}h{\partial _{y}}h+\eta .
\label{cross-gen}
\end{equation}%
Indeed, it is easy to check that the probability density 
\begin{equation}
\mathcal{P}=N\exp \left\{ -\frac{1}{D}\int dxdy\left[ F_{x}({\partial _{x}}%
h)+F_{y}({\partial _{y}}h)\right] \right\} ,
\end{equation}%
where $f_{x}(u)=dF_{x}(u)/du$ and $f_{y}(u)=dF_{y}(u)/du$, is stationary.

\section{General Prescriptions for Gaussian Steady States}

\label{general}

The solution of Fokker--Planck equations by direct check of Ans\"{a}tze soon
becomes laborious beyond simple one and two dimensional examples. Instead,
we derive general prescriptions on the structure of the equations of
evolution, for the realization of Gaussian steady states. We consider
equations of the form 
\begin{equation}
\partial _{t}h_{i}(\mathbf{x},t)=\mathcal{L}_{\mathbf{x}}^{(i)}[h]+\mathcal{N%
}_{\mathbf{x}}^{(i)}[h]+\eta _{i}(\mathbf{x},t),
\end{equation}%
for $n$ coupled fields ($i=1,\ldots ,n$), $\mathbf{x}\in \mathbf{R}^{d}$,
and Gaussian (thermal) noise with correlator 
\begin{equation}
\langle \eta _{i}(\mathbf{x},t)\eta _{j}(\mathbf{x}^{\prime },t^{\prime
})\rangle =2D_{i}\delta _{ij}\delta ^{d}(\mathbf{x}-\mathbf{x}^{\prime
})\delta (t-t^{\prime }).
\end{equation}%
$\mathcal{L}_{\mathbf{x}}$ and $\mathcal{N}_{\mathbf{x}}$ denote linear and
non-linear functionals of the fields, respectively, evaluated at point $%
\mathbf{x}$. If $\eta _{i}$ and $\eta _{j\neq i}$ are uncorrelated \cite%
{diagonalize}, the Fokker--Planck equation reads 
\begin{equation}
\partial _{t}\mathcal{P}=\int d^{d}x\sum_{i}\frac{\delta }{\delta h_{i}}%
\left[ -(\mathcal{L}_{\mathbf{x}}^{(i)}+\mathcal{N}_{\mathbf{x}}^{(i)})%
\mathcal{P}+\frac{\delta \mathcal{P}}{\delta h_{i}}\right] ,
\end{equation}%
where we have absorbed $D_{i}$ in a rescaling of $h_{i}$ (by $\sqrt{D_{i}}$%
), and reduces to 
\begin{equation}
\partial _{t}\mathcal{P}=\int d^{d}x\sum_{i}\left[ -(\mathcal{L}_{\mathbf{x}%
}^{(i)}+\mathcal{N}_{\mathbf{x}}^{(i)})\frac{\delta \mathcal{P}}{\delta h_{i}%
}+\frac{\delta ^{2}\mathcal{P}}{\delta h_{i}^{2}}\right] 
\end{equation}%
if $\mathcal{L}$ and $\mathcal{N}$ depend upon the derivatives of $h$ only.
We are looking for a Gaussian probability density 
\begin{equation}
\mathcal{P}=Ne^{-Q[h]},  \label{q}
\end{equation}%
with $Q[h]$ a quadratic form and $N$ a normalizing factor, that solves the
steady state $\partial _{t}\mathcal{P}=0$, \textit{i.e.} 
\begin{equation}
\int d^{d}x\sum_{i}\left[ -(\mathcal{L}_{\mathbf{x}}^{(i)}+\mathcal{N}_{%
\mathbf{x}}^{(i)})\left( -\frac{\delta Q}{\delta h_{i}}\right) +\left( \frac{%
\delta Q}{\delta h_{i}}\right) ^{2}\right] =0.
\end{equation}%
The quadratic terms cancel if $\mathcal{L}$ and $Q$ are related through 
\begin{equation}
\mathcal{L}_{\mathbf{x}}^{(i)}=-\frac{\delta Q}{\delta h_{i}},
\label{relating}
\end{equation}%
and it remains to find the form of $\mathcal{L}$ and $\mathcal{N}$ for which
the integral%
\begin{eqnarray}
\mathcal{J}[h] &\equiv &\int d^{d}x\sum_{i}\mathcal{N}_{\mathbf{x}}^{(i)}%
\frac{\delta Q}{\delta h_{i}}  \nonumber \\
&=&-\int d^{d}x\sum_{i}\mathcal{N}_{\mathbf{x}}^{(i)}\mathcal{L}_{\mathbf{x}%
}^{(i)}  \label{(i)}
\end{eqnarray}%
vanishes. This is the case if the integrand either vanishes identically or
is the divergence of a vector field. In either case, the integrand is
unchanged by a variation $\delta h(\mathbf{x})$ that vanishes at infinity.
If $\delta h$ is localized at $\mathbf{x}$, the condition $\mathcal{J}%
[h+\delta h]=J[h]$ translates, to first order in $\delta h$, into 
\begin{equation}
\frac{\delta }{\delta h_{j}(\mathbf{x})}\int d^{d}y\mathcal{N}_{\mathbf{y}%
}^{(i)}\mathcal{L}_{\mathbf{y}}^{(i)}=0,  \label{variation}
\end{equation}%
where the summation over $i$ is understood.

For coupled KPZ-like equations \cite{drift} of the form 
\begin{equation}
\partial _{t}h_{i}=\nu _{ij\alpha \beta }\partial _{\alpha }\partial _{\beta
}h+\frac{1}{2}\lambda _{ijk\alpha \beta }\partial _{\alpha }h_{j}\partial
_{\beta }h_{k}+\eta _{i},  \label{genKPZ}
\end{equation}%
where Latin indices $i,j,k=1,\ldots ,n$ refer to field components and Greek
indices $\alpha ,\beta =1,\ldots ,d$ refer to spatial components, Eq.~(\ref%
{variation}) reduces to%
\begin{equation}
\begin{array}{l}
\nu _{ij\alpha \beta }\lambda _{ikl\gamma \delta }(\partial _{\alpha
}\partial _{\beta }\partial _{\gamma }h_{k}\partial _{\delta }h_{l}+\partial
_{\alpha }\partial _{\gamma }h_{k}\partial _{\beta }\partial _{\delta }h_{l})
\\ 
-\nu _{ik\alpha \beta }\lambda _{ijl\gamma \delta }(\partial _{\alpha
}\partial _{\beta }\partial _{\gamma }h_{k}\partial _{\delta }h_{l}+\partial
_{\alpha }\partial _{\beta }h_{k}\partial _{\gamma }\partial _{\delta
}h_{l})=0,%
\end{array}
\label{cond}
\end{equation}%
after some renaming of the indices and using the fact that $\nu $ and $%
\lambda $ can always be chosen to satisfy the equalities 
\begin{equation}
\nu _{ij\alpha \beta }=\nu _{ij\beta \alpha }\quad \mathrm{and}\quad \lambda
_{ijk\alpha \beta }=\lambda _{ikj\beta \alpha }.
\end{equation}%
Furthermore, Eq.~(\ref{relating}) requires the symmetry 
\begin{equation}
\nu _{ij\alpha \beta }=\nu _{ji\alpha \beta },
\end{equation}%
and the stationary probability density reads 
\begin{equation}
\mathcal{P}=N\exp \left( -\int d^{d}x\frac{\nu _{ij\alpha \beta }}{2}%
\partial _{\alpha }h_{i}\partial _{\beta }h_{j}\right) .
\end{equation}%
Since repeated indices are summed over, Eq.~(\ref{cond}) represents $n$
conditions---one for each possible value of the index $j$. Each of these in
fact encapsulates more than one constraint: Eq.~(\ref{cond}) is composed of
terms that come in one of two derivative structures and, as the equation
must be true for arbitrary $h$, terms of a given derivative structure must
cancel independently. This is ensured by the following two sets of \textit{\
generalized fluctuation--dissipation conditions} on the tensors $\nu $ and $%
\lambda $: The first condition comes from grouping terms in Eq.~(\ref{cond})
which are products of first and third derivatives, (such as $\partial
_{\alpha }\partial _{\beta }\partial _{\gamma }h_{k}\partial _{\delta }h_{l}$%
), and reads 
\begin{equation}
\begin{array}{lll}
\text{(I)\qquad }\sum_{P}( & \nu _{ijP(\alpha )P(\beta )}\lambda
_{iklP(\gamma )\delta } &  \\ 
& -\nu _{ikP(\alpha )P(\beta )}\lambda _{ijlP(\gamma )\delta } & )=0,%
\end{array}%
\end{equation}%
where the summation runs over the six permutations of the indices $\alpha
,\beta ,\gamma $. A second condition comes from grouping terms of the form $%
\partial _{\alpha }\partial _{\gamma }h_{k}\partial _{\beta }\partial
_{\delta }h_{l}$, and gives 
\begin{equation}
\begin{array}{lll}
\text{(II)\qquad }\sum_{R,R^{\prime }}( & 2\nu _{ijR(\alpha )R^{\prime
}(\gamma )}\lambda _{iklR(\beta )R^{\prime }(\delta )} &  \\ 
& -\nu _{ikR(\alpha )R(\beta )}\lambda _{ijlR^{\prime }(\gamma )R^{\prime
}(\delta )} &  \\ 
& -\nu _{ilR^{\prime }(\gamma )R^{\prime }(\delta )}\lambda _{ijkR(\alpha
)R(\beta )} & )=0,%
\end{array}%
\end{equation}%
where the summation runs over the two permutations of the indices $\alpha
,\beta $ and the two permutations of the indices $\gamma ,\delta $. Each
pair of conditions corresponds to a given choice of numerical values for $j$%
, $k$, $l$, $\alpha $, $\beta $, $\gamma $, and $\delta $.

Conditions (I) and (II) are \textit{necessary} for a Gaussian solution of
the steady-state Fokker--Planck equation. They are also \textit{sufficient}
conditions, since $\mathcal{J}[h+\delta h]=\mathcal{J}[h]$ to first order 
\textit{for any} $\delta h$ implies $\mathcal{J}[h+\delta h]=\mathcal{J}[h]$
to all orders, and $\mathcal{J}[h]$~=~constant~$\equiv c$. But normalization
of $\mathcal{P}$ allows only $c=0$ (otherwise $\mathcal{P}$ either increases
or decreases uniformly), and consequently $\partial _{t}\mathcal{P}=0$.

\section{Absence of Gaussian Steady States in Three and Higher Dimensions}

\label{3d}

If the matrix of Laplacian coefficients $\nu _{ij\alpha \beta }$ is positive
definite, as required by infra-red stability, it can be diagonalized into 
\begin{equation}
\nu _{ij\alpha \beta }=\nu _{ij}\delta _{\alpha \beta },  \label{diagonal}
\end{equation}%
by successive rotations and rescalings, and the prescriptions (I) and (II)
simplify correspondingly. In three and higher dimensions we may choose the
space indices, in applying the prescriptions, such that $\alpha =\gamma $
while $\alpha \not=\beta \not=\delta \not=\alpha $. With this choice, it is
straightforward to check that prescription (II) forbids a non-vanishing
contraction of tensors with different space indices, hence enforces the
partially diagonal form 
\begin{equation}
\nu _{ij}\lambda _{ikl\alpha \beta }\equiv \nu _{ij}\lambda _{ikl\alpha
\alpha }\delta _{\alpha \beta },
\end{equation}%
(where $i$ is summed over but $\alpha $ is not). With this constraint,
prescription (II) takes the form 
\begin{equation}
\begin{array}{lll}
\sum_{R,R^{\prime }}[ & \nu _{ij}\lambda _{jklR(\beta )R(\beta )}\delta
_{R(\alpha )R^{\prime }(\gamma )}\delta _{R(\beta )R^{\prime }(\delta )} & 
\\ 
& -\frac{1}{2}(\nu _{ik}\lambda _{ijlR^{\prime }(\gamma )R^{\prime }(\gamma
)}+\nu _{il}\lambda _{ijkR(\alpha )R(\alpha )}) &  \\ 
& \times \delta _{R(\alpha )R(\beta )}\delta _{R^{\prime }(\gamma )R^{\prime
}(\delta )} & ]=0.%
\end{array}
\label{cond-special}
\end{equation}%
This equation expresses a set of different conditions, one for each choice
of values of the indices that are not summed over. Specifically, for a
particular choice in which $\alpha =\gamma \not=\beta =\delta $, Eq. (\ref%
{cond-special}) translates into 
\begin{equation}
\nu _{ij}\lambda _{ikl\alpha \alpha }=-\nu _{ij}\lambda _{ikl\beta \beta }.
\end{equation}%
For the sake of visual ease, let us define the object $\varphi _{\alpha
}\equiv \nu _{ij}\lambda _{ikl\alpha \alpha }$ (the dependence of $\varphi
_{\alpha }$ on $j,k,l$ is tacit), in terms of which this identity reads%
\begin{equation}
\varphi _{\alpha }=-\varphi _{\beta };
\end{equation}%
clearly, if it is possible to choose three or more distinct values of the
indices $\alpha ,\beta $, this condition is frustrated and admits only the
trivial solution%
\begin{equation}
\varphi _{\alpha }=\nu _{ij}\lambda _{ikl\alpha \alpha }=0  \label{phi}
\end{equation}%
for all $j,k,l,\alpha $. Viewed as a vector identity, this requires that any
vector $\mathbf{\nu }_{j}$ be orthogonal to any vector $\mathbf{\lambda }%
_{kl\alpha \alpha }$ (with components labeled by $i=1,...n$). As long as the
matrix $\nu _{ij}$ is non-degenerate, there are $n$ non-vanishing
independent vectors $\mathbf{\nu }_{j}$ (for $j=1,...n$) and Eq.~(\ref{phi})
is satisfied only if the vectors $\mathbf{\lambda }_{kl\alpha \alpha }$
vanish for all $k,l,\alpha $. Hence, no Gaussian steady state is achievable
in three and higher dimensions if non-linearities are present in the
equations of evolution \cite{Tauber}.

\section{Coupled Fields in One and Two Dimensions}

\label{coupled}

In the case of a single field fluctuating in two dimensions, prescriptions
(I) and (II) immediately enforce the form of Eq.~(\ref{antisym-gen}) for
which a Gaussian steady state may be reached, as we checked explicitely in
Sec. \ref{2d}. In what follows, we discuss examples in which the coupling
among several fluctuating fields broadens the class of non-linear equations
with Gaussian steady states beyond this specific anisotropic form (with
coefficients of opposite signs).

\subsection{Coupled Lines and Drifting Polymers}

\label{lines}

An array of fluctuating directed lines \cite{HHF,KZ,FH} is parametrized by a
single variable, consequently the Greek indices in Eq.~(\ref{genKPZ}) all
take the same value and may be omitted. Equation~(\ref{genKPZ}) then reduces
to 
\begin{equation}
\partial _{t}h_{i}=\nu _{ij}{\partial _{x}}{\partial _{x}}h_{j}+\frac{1}{2}%
\lambda _{ijk}{\partial _{x}}h_{j}{\partial _{x}}h_{k}+\eta _{i},
\label{genKPZ-1d}
\end{equation}%
a generalization of the one dimensional KPZ equation for several coupled
fields. In this simple case, prescriptions (I) and (II) are fulfilled by any
tensors $\nu $, $\lambda $ such that 
\begin{equation}
\nu _{ij}\lambda _{ikl}=\nu _{ik}\lambda _{ijl}=\nu _{il}\lambda _{ikj},
\label{cond-1d}
\end{equation}%
where $\lambda _{ijk}$ can always be chosen symmetric in $j,k$, and the sum
over $i$ is understood. It is easy to check that these relations ensure a
stationary probability density 
\begin{equation}
\mathcal{P}=N\exp \left( -\int dx\frac{\nu _{ij}}{2}{\partial _{x}}h_{i}{%
\partial _{x}}h_{j}\right) .
\end{equation}%
Indeed, with this Ansatz%
\begin{eqnarray}
\partial _{t}\mathcal{P} &=&-\int dx\frac{1}{2}\lambda _{ikl}{\partial _{x}}%
h_{k}{\partial _{x}}h_{l}\nu _{ij}{\partial _{x}}{\partial _{x}}h_{j}%
\mathcal{P} \\
&=&-\int dx\frac{1}{6}(\nu _{ij}\lambda _{ikl}{\partial _{x}}{\partial _{x}}%
h_{j}{\partial _{x}}h_{k}{\partial _{x}}h_{l}  \nonumber \\
&&+\nu _{ik}\lambda _{ijl}{\partial _{x}}h_{j}{\partial _{x}}{\partial _{x}}%
h_{k}{\partial _{x}}h_{l}  \nonumber \\
&&+\nu _{il}\lambda _{ikj}{\partial _{x}}h_{j}{\partial _{x}}h_{k}{\partial
_{x}}{\partial _{x}}h_{l})\mathcal{P},  \nonumber
\end{eqnarray}%
where we have renamed the mute indices to obtain the last equality. But Eq. (%
\ref{cond-1d}) implies that all three coefficients are identical, and 
\begin{equation}
\partial _{t}\mathcal{P}=-\int dx\frac{1}{6}\nu _{ij}\lambda _{ikl}{\partial
_{x}}({\partial _{x}}h_{j}{\partial _{x}}h_{k}{\partial _{x}}h_{l})=0.
\end{equation}

A special case of Eq.~(\ref{genKPZ-1d}), 
\begin{equation}
\left\{ 
\begin{array}{l}
\partial _{t}h_{\Vert }=\nu _{\Vert }{\partial _{x}}{\partial _{x}}h_{\Vert
}+\frac{1}{2}\lambda _{\Vert }({\partial _{x}}h_{\Vert })^{2}+\frac{1}{2}%
\lambda _{\perp }({\partial _{x}}h_{\perp })^{2}+\eta _{\Vert } \\ 
\partial _{t}h_{\perp }=\nu _{\perp }{\partial _{x}}{\partial _{x}}h_{\perp
}+\lambda _{\times }{\partial _{x}}h_{\Vert }{\partial _{x}}h_{\perp }+\eta
_{\perp }%
\end{array}%
,\right.  \label{polymer}
\end{equation}%
was introduced in Refs. \cite{Ertas} to describe a directed polymer drifting
perpendicularly to itself. Here $h_{\Vert }$ and $h_{\perp }$ are
interpreted not as height fields associated with two different lines
embedded in two dimensions, but rather as dynamically coupled longitudinal
and transverse (to the average velocity of the polymer) fluctuations of a
single line embedded in three dimensions. The conditions of Eq.~(\ref%
{cond-1d}) for a Gaussian steady state simplify to 
\begin{equation}
\nu _{\Vert }\lambda _{\perp }=\nu _{\perp }\lambda _{\times },
\label{fol-polymer}
\end{equation}%
in agreement with a direct check \cite{Ertas,temperature}.

In the simplest case with identical longitudinal and transverse
coefficients, a stationary Gaussian distribution follows trivially from the
steady-state properties of the one-dimensional KPZ equation (discussed in
Sec. \ref{1d}), since the equations 
\begin{equation}
\left\{ 
\begin{array}{l}
\partial _{t}h_{\parallel }=\nu {\partial _{x}}\partial _{x}h_{\parallel }+%
\frac{1}{2}\lambda \lbrack (\partial _{x}h_{\parallel })^{2}+(\partial
_{x}h_{\perp })^{2}]+\eta _{\parallel } \\ 
\partial _{t}h_{\perp }=\nu \partial _{x}\partial _{x}h_{\perp }+\lambda
\partial _{x}h_{\parallel }\partial _{x}h_{\perp }+\eta _{\perp }%
\end{array}%
\right.
\end{equation}%
are equivalent to 
\begin{equation}
\left\{ 
\begin{array}{c}
\partial _{t}h_{+}=\nu \partial _{x}\partial _{x}h_{+}+\frac{1}{2}\lambda
(\partial _{x}h_{+})^{2}+\eta _{+} \\ 
\partial _{t}h_{-}=\nu \partial _{x}\partial _{x}h_{-}+\frac{1}{2}\lambda
(\partial _{x}h_{-})^{2}+\eta _{-}%
\end{array}%
,\right.
\end{equation}%
with $h_{\pm }=h_{\parallel }\pm h_{\perp }$ and $\eta _{\pm }=\eta
_{\parallel }\pm \eta _{\perp }$. Clearly, the remark extends to higher
spatial dimensions, where 
\begin{equation}
\left\{ 
\begin{array}{l}
\partial _{t}h_{\parallel }=\nu \nabla ^{2}h_{\parallel }+\frac{1}{2}\lambda
\lbrack (\nabla h_{\parallel })^{2}+(\nabla h_{\perp })^{2}]+\eta
_{\parallel } \\ 
\partial _{t}h_{\perp }=\nu \nabla ^{2}h_{\perp }+\lambda \nabla
h_{\parallel }\cdot \nabla h_{\perp }+\eta _{\perp }%
\end{array}%
,\right.  \label{special}
\end{equation}%
transform into 
\begin{equation}
\left\{ 
\begin{array}{l}
\partial _{t}h_{+}=\nu \nabla ^{2}h+\frac{1}{2}\lambda (\nabla
h_{+})^{2}+\eta _{+} \\ 
\partial _{t}h_{+}=\nu \nabla ^{2}h_{-}+\frac{1}{2}\lambda (\nabla
h_{-})^{2}+\eta _{-}%
\end{array}%
.\right.
\end{equation}%
Therefore, the \textquotedblleft roughness\textquotedblright\ exponent $%
\zeta _{\perp }$ of a passive scalar $h_{\perp }$ advected \cite%
{Kraichman,Turbulence} by a Burgers flow $\mathbf{u}=\nabla h_{\parallel }$,
defined through 
\begin{equation}
\left\langle \left( h_{\perp }(\mathbf{x})-h_{\perp }(\mathbf{y})\right)
^{2}\right\rangle ^{1/2}\sim |\mathbf{x}-\mathbf{y}|^{\zeta _{\perp }},
\end{equation}%
is none other than the KPZ roughness exponent \cite{KPZ}.

\subsection{Magnetic Growth}

\label{magnetic}

In a growing magnetic material, the spins may be assumed frozen in the bulk
while still fluctuating on the surface, which itself fluctuates in height 
\cite{Drossel,DK}. For the case of XY spins, described by a single angular
field $\theta (x,y,t)$, Ref. \cite{Kardar99} notes that a two-dimensional
version of Eqs.~(\ref{polymer}) governs these non-equilibrium coupled
fluctuations. In the modified notation, these equations of evolution read 
\begin{equation}
\left\{ 
\begin{array}{l}
\partial _{t}h=\nu _{h}\nabla ^{2}h+\frac{1}{2}\lambda _{hh}(\nabla h)^{2}+%
\frac{1}{2}\lambda _{\theta \theta }(\nabla \theta )^{2}+\eta _{h} \\ 
\partial _{t}\theta =\nu _{\theta }\nabla ^{2}\theta +\lambda _{h\theta
}\nabla h\cdot \nabla \theta +\eta _{\theta }%
\end{array}%
\right. .  \label{xy}
\end{equation}%
(Obviously, Eqs.~(\ref{xy}) fail to capture the periodic nature of $\theta $%
, and with it the potentially relevant presence of topological defects.)

Using prescriptions (I) and (II), it is easy to show that no Gaussian steady
state exists for such isotropic equations as long as $h$ and $\theta $ are
decoupled at the linear level. However, the stationary Gaussian distribution 
\begin{equation}
\mathcal{P}=N\exp \left( -\int dxdy\left[ \frac{\nu _{h}}{2}(\nabla h)^{2}+%
\frac{\nu _{\theta }}{2}(\nabla \theta )^{2}\right] \right),
\end{equation}%
is achieved by a natural extension of the anisotropic Eq.~(\ref{antisym}), 
\begin{equation}
\left\{ 
\begin{array}{lll}
\partial _{t}h & = & \nu _{h}\nabla ^{2}h+\frac{1}{2}\lambda _{hh}[({%
\partial _{x}}h)^{2}-({\partial _{y}}h)^{2}] \\ 
&  & +\frac{1}{2}\lambda _{\theta \theta }[({\partial _{x}}\theta )^{2}-({%
\partial _{y}}\theta )^{2}]+\eta _{h} \\ 
\partial _{t}\theta & = & \nu _{\theta }\nabla ^{2}\theta +\lambda _{h\theta
}({\partial _{x}}h{\partial _{x}}\theta -{\partial _{y}}h{\partial _{y}}%
\theta )+\eta _{\theta }%
\end{array}%
,\right.  \label{vw-htheta}
\end{equation}%
provided $\nu _{h}\lambda _{\theta \theta }=\nu _{\theta }\lambda _{h\theta
} $ (in direct analogy to Eq.~(\ref{fol-polymer})). We note also that these
equations again satisfy the symmetry of Eq.~(\ref{symmetry}).

\subsection{Crystalline Growth}

\label{Xtal}

The height fluctuations of a material characterized by internal order
parameters, as in the above case of a growing XY magnet, are subjected to
the fluctuations of these order parameters. Conversely, the evolution of the
internal order parameters depends on the height fluctuations. In contrast to
amorphous growth, we may say that the fluctuations of an ordered material
results from a \textit{textural growth}, as the the additional fields invest
the interface with a texture that constrains its fluctuations. The prime
example is that of the growth of a crystal in which surface phonons interact
with height fluctuations. In analogy with Eqs.~(\ref{xy}), we propose the
following equations for isotropic crystalline growth,%
\begin{equation}
\left\{ 
\begin{array}{lll}
\partial _{t}h & = & \nu _{h}\nabla ^{2}h+\frac{1}{2}\lambda _{hh}(\nabla
h)^{2}+\frac{1}{2}\lambda _{uu}^{(1)}(\nabla \cdot \mathbf{u})^{2} \\ 
&  & +\frac{1}{2}\lambda _{uu}^{(2)}\nabla u_{i}\cdot \nabla u_{i}+\frac{1}{2%
}\lambda _{uu}^{(3)}\partial _{i}\mathbf{u}\cdot \nabla u_{i}+\eta _{h} \\ 
\partial _{t}u_{i} & = & \nu _{u}^{(1)}\nabla ^{2}u_{i}+\nu
_{u}^{(2)}\partial _{i}\nabla \cdot \mathbf{u}+\lambda _{hu}^{(1)}\partial
_{i}h\nabla \cdot \mathbf{u} \\ 
&  & +\lambda _{hu}^{(2)}\nabla h\cdot \nabla u_{i}+\lambda
_{hu}^{(3)}\nabla h\cdot \partial _{i}\mathbf{u}+\eta _{i}%
\end{array}%
,\right.  \label{crystal}
\end{equation}%
where $\mathbf{u}(x,y,t)$ is the surface displacement vector field, and $\nu
_{u}^{(1)}$, $\nu _{u}^{(2)}$ are related to the usual Lam\'{e} coefficients
through 
\begin{equation}
\left\{ 
\begin{array}{l}
\nu _{u}^{(1)}=\mu _{\mathrm{Lam\acute{e}}}\nonumber \\ 
\nu _{u}^{(2)}=\mu _{\mathrm{Lam\acute{e}}}+\lambda _{\mathrm{Lam\acute{e}}}%
\end{array}%
.\right.
\end{equation}%
One would like to know, given the richness of Eqs.~(\ref{crystal}), what
phases they describe beyond the usual KPZ (amorphous) phase. As a first step
towards a complete answer, we discuss crystalline growth equations that
admit an exact Gaussian steady state.

Equations (\ref{crystal}) tacitly assume a triangular lattice at the
microscopic level, since other lattices would be reflected in an anisotropic
continuum limit that reproduces the appropriate crystal symmetries. In the
context of anistropic equations, trivial extensions of Eqs.~(\ref{vw-htheta}%
) support a Gaussian steady state.

Rather than dwelling on these examples, we turn to a new possibility, namely
isotropic equations of the form of Eqs.~(\ref{crystal}) that admit a
stationary Gaussian probability density, as allowed by the presence of
linear couplings between $u_{x}$ and $u_{y}$. A somewhat tedious but
straightforward application of prescriptions (I) and (II) to Eqs.~(\ref%
{crystal}) yields two (and only two) non-trivial solutions, characterized by
a vanishing shear modulus $\nu _{u}^{(1)}=0$ and a vanishing bulk modulus $%
\nu _{u}^{(1)}+\nu _{u}^{(2)}=0$.

With a vanishing shear modulus $\nu _{u}^{(1)}=0$, (I) and (II) imply 
\begin{equation}
\lambda _{hh}=\lambda _{uu}^{(2)}=\lambda _{uu}^{(3)}=\lambda
_{hu}^{(2)}=\lambda _{hu}^{(3)}=0
\end{equation}%
and%
\begin{equation}
\nu _{h}\lambda _{uu}^{(1)}=\nu _{u}^{(2)}\lambda _{hu}^{(1)},
\end{equation}%
resulting in the equations of motion%
\begin{equation}
\left\{ 
\begin{array}{lll}
\partial _{t}h & = & \nu _{h}\nabla ^{2}h+\frac{1}{2}\lambda _{uu}(\nabla
\cdot \mathbf{u})^{2}+\eta _{h} \\ 
\partial _{t}\mathbf{u} & = & \nu _{u}\nabla (\nabla \cdot \mathbf{u}%
)+\lambda _{hu}\nabla h(\nabla \cdot \mathbf{u})+\mathbf{\eta }%
\end{array}%
,\right.  \label{shear}
\end{equation}%
where the coefficients have been renamed ($\nu _{u}\equiv \nu _{u}^{(2)}$, $%
\lambda _{uu}\equiv \lambda _{uu}^{(1)}$, $\lambda _{hu}\equiv \lambda
_{hu}^{(1)}$). The probability density 
\begin{equation}
\mathcal{P}=N\exp \left( -\int dxdy\left[ \frac{\nu _{u}}{2}(\nabla h)^{2}+%
\frac{\nu _{u}}{2}(\nabla \cdot \mathbf{u})^{2}\right] \right)
\end{equation}%
is stationary, as the direct check 
\begin{eqnarray}
\partial _{t}\mathcal{P} &=&\int dxdy\left[ \frac{1}{2}\lambda _{uu}(\nabla
\cdot \mathbf{u})\nu _{h}\nabla ^{2}h\right. \\
&&\left. +\frac{\nu _{h}}{\nu _{u}}\lambda _{uu}\nabla h(\nabla \cdot 
\mathbf{u})\nu _{u}\nabla (\nabla \cdot \mathbf{u})\right]  \nonumber \\
&=&-\int dxdy\frac{\nu _{h}\lambda _{uu}}{2}\nabla \cdot \lbrack \nabla
h(\nabla \cdot \mathbf{u})^{2}] \\
&=&0,
\end{eqnarray}%
confirms. In terms of the density fluctuations 
\begin{equation}
\rho \equiv \nabla \cdot u
\end{equation}%
and the vorticity 
\begin{equation}
\Omega \equiv {\partial _{x}}u_{y}-{\partial _{y}}u_{x},
\end{equation}%
we may interpret Eqs.~(\ref{shear}) as describing a liquid that is flowing
on a fluctuating surface $h(x,y,t)$ via the partially decoupled set of
equations 
\begin{equation}
\left\{ 
\begin{array}{l}
\partial _{t}h=\nu _{h}\nabla ^{2}h+\frac{1}{2}\lambda _{uu}\rho ^{2}+\eta
_{h}\nonumber \\ 
\partial _{t}\rho =\nu _{u}\nabla ^{2}\rho +\lambda _{hu}\nabla \cdot (\rho
\nabla h)+\nabla \cdot \mathbf{\eta }\nonumber \\ 
\partial _{t}\Omega ={\partial _{x}}\eta _{y}-{\partial _{y}}\eta _{x}%
\end{array}%
\right.  \label{density}
\end{equation}%
This is again reminiscent of the advection of a scalar \cite%
{Kraichman,Turbulence} (density fluctuations are advected along height
gradients) which is not quite passive, as $\rho$ influences the evolution of 
$h$. Equations (\ref{density}) describe also a special case of the coupled
growth of a binary film or, equivalently, of a scalar (Ising) magnet \cite%
{DK}.

With a vanishing bulk modulus $\nu _{u}^{(1)}+\nu _{u}^{(2)}=0$, (I) and
(II) imply 
\begin{equation}
\lambda _{hh}=\lambda _{uu}^{(1)}=\lambda _{hu}^{(1)}=\lambda
_{uu}^{(2)}+\lambda _{uu}^{(3)}=\lambda _{hu}^{(2)}+\lambda _{hu}^{(3)}=0
\end{equation}%
and%
\begin{equation}
\nu _{h}\lambda _{uu}^{(2)}=\nu _{u}^{(1)}\lambda _{hu}^{(2)},
\end{equation}%
resulting in the equations of motion%
\begin{equation}
\left\{ 
\begin{array}{lll}
\partial _{t}h & = & \nu _{h}\nabla ^{2}h+\frac{1}{2}\lambda _{uu}\partial
_{i}u_{j}\left( \partial _{i}u_{j}-\partial _{j}u_{i}\right) +\eta _{h} \\ 
\partial _{t}u_{i} & = & \nu _{u}\partial _{j}\left( \partial
_{j}u_{i}-\partial _{i}u_{j}\right)  \\ 
&  & +\lambda _{hu}\partial _{j}h\left( \partial _{j}u_{i}-\partial
_{i}u_{j}\right) +\eta 
\end{array}%
,\right.   \label{bulk}
\end{equation}%
where $\nu _{u}\equiv \nu _{u}^{(1)}$, $\lambda _{uu}\equiv \lambda
_{uu}^{(2)},$ and $\lambda _{hu}\equiv \lambda _{hu}^{(2)}$. Introducing a
fictitious direction $z$ according to the natural definition $\nabla \times 
\mathbf{u}=(\partial _{x}u_{y}-\partial _{y}u_{x})\hat{z}$ with $\nabla
=(\partial _{x},\partial _{y},0)$ and $\mathbf{u}=(u_{x},u_{y},0)$, we can
rewrite Eqs.~(\ref{bulk}) in the more compact form 
\begin{equation}
\left\{ 
\begin{array}{lll}
\partial _{t}h & = & \nu _{h}\nabla ^{2}h+\frac{1}{2}\lambda _{uu}(\nabla
\times \mathbf{u})^{2}+\eta _{h} \\ 
\partial _{t}\mathbf{u} & = & -\nu _{u}\nabla \times (\nabla \times \mathbf{u%
}) \\ 
&  & -\lambda _{hu}\nabla h\times (\nabla \times \mathbf{u})+\mathbf{\eta }%
\end{array}%
\right. .  \label{bulk2}
\end{equation}%
With this form at hand, the stationarity of the probability density 
\begin{equation}
\mathcal{P}=N\mathrm{exp}\left( -\int dxdy\left[ \frac{\nu _{h}}{2}(\nabla
h)^{2}+\frac{\nu _{u}}{2}(\nabla \times \mathbf{u})^{2}\right] \right) 
\end{equation}%
is obtained in direct analogy with the case of vanishing shear modulus. In
terms of $\rho $ and $\Omega $, Eqs.~(\ref{bulk2}) become%
\begin{equation}
\left\{ 
\begin{array}{lll}
\partial _{t}h & = & \nu _{h}\nabla ^{2}h+\frac{1}{2}\lambda _{uu}\Omega
^{2}+\eta _{h} \\ 
\partial _{t}\Omega  & = & \nu _{u}\nabla ^{2}\Omega +\lambda _{hu}\nabla
\cdot (\Omega \nabla h)+\eta _{\perp } \\ 
\partial _{t}\rho  & = & \eta _{\Vert }%
\end{array}%
\right. ,  \label{dislocations}
\end{equation}%
none other than Eqs.~(\ref{density}) with $\rho $ and $\Omega $
interchanged, but still a transverse noise $\eta _{\perp }=\partial _{x}\eta
_{y}-{\partial _{y}}\eta _{x}$ driving the fluctuations of $\Omega $ and a
longitudinal noise $\eta _{\Vert }={\partial _{x}}\eta _{x}+{\partial _{y}}%
\eta _{y}$ driving $\rho $. Here we can interpret $\Omega $ as the density
of dislocations whose presence locally affects growth.

Experiments on surface growth do not point to a unique characterization \cite%
{Krim} and, in particular, most measured roughness exponents differ from
those expected on the basis of the KPZ equation \cite{Reviews,Family,Meakin}%
. This may result from conservation laws, incorporated in some molecular
beam epitaxy models \cite{Sun,Lai,Siegert,Krug}, but it also may be the
consequence of the dynamic coupling of the height fluctuations with the
fluctuations of the intrinsic order parameter of the material, as in
magnetic or crystalline growth. Thus, having established that the special
Eqs.~(\ref{shear}) and (\ref{bulk2}) admit Gaussian steady states, one would
like to know whether the full Eqs. (\ref{crystal}) may flow to them under
renormalization. If Eqs.~(\ref{shear}) or (\ref{bulk2}) indeed have a basin
of attraction, we are left with the surprising conclusion that the coupling
to crystal vibration tethers the fluctuating surface, in an appropriately
prepared sample, to logarithmic roughness, in marked contrast to the KPZ
roughness associated with amorphous growth.

Rather than renormalizing the full set of Eqs.~(\ref{crystal}), one may, as
a first attempt, consider the renormalization of Eqs.~(\ref{density}) or (%
\ref{dislocations}) with an added KPZ term $(\lambda _{hh}\not=0)$, 
\begin{equation}
\left\{ 
\begin{array}{lll}
\partial _{t}h & = & \nu _{h}\nabla ^{2}h+\frac{1}{2}\lambda _{hh}(\nabla
h)^{2}+\frac{1}{2}\lambda _{uu}\rho ^{2}+\eta _{h} \\ 
\partial _{t}\rho & = & \nu _{u}\nabla ^{2}\rho +\lambda _{hu}\nabla \cdot
(\rho \nabla h)+\eta _{\Vert } \\ 
\partial _{t}\Omega & = & \eta _{\perp }%
\end{array}%
\right. .  \label{added}
\end{equation}%
The sub-space of Eqs.~(\ref{added}) is closed under renormalization since
they are invariant under the transformation 
\begin{equation}
\mathbf{u}\rightarrow \mathbf{u}+\nabla \times \mathbf{A},
\end{equation}%
with $\mathbf{A}(x,y)$ an arbitrary smooth function. (Similarly, Eqs.~(\ref%
{dislocations}) are invariant under 
\begin{equation}
\mathbf{u}\rightarrow \mathbf{u}+\nabla \phi ,
\end{equation}%
with $\phi $ arbitrary.) Also, Eqs.~(\ref{added}) are interesting in their
own right: the field $\rho $ may be interpreted as density fluctuations, 
\textit{e.g.} of surfactants, sliding on the surface. Similarly, if $h$
describes the height of a material surface (such as a liquid film), $\rho $
may play the role of material surface density fluctuations (proportional to
the thickness of the film in the case of an incompressible liquid). These
density fluctuations are conserved as the particles, or liquid elements,
slide along gradients in the surface height, according to Eqs. (\ref{added}%
). However, the implicit $\rho \rightarrow -\rho $ symmetry of Eqs. (\ref%
{added}) is difficult to justify physically for surfactants or liquid films.

As a final non-perturbative observation on Eq.~(\ref{added}), we note that
they are invariant under the infinitesimal tilt 
\begin{equation}
\left\{ 
\begin{array}{l}
\mathbf{x}\rightarrow \mathbf{x}+\lambda \mathbf{\epsilon }t\nonumber \\ 
h\rightarrow h+\mathbf{\epsilon }\cdot \mathbf{x}%
\end{array}%
\right.
\end{equation}%
in the sub-space $\lambda _{hh}=\lambda _{hu}\equiv \lambda $, so that $%
\lambda _{hh}$ and $\lambda _{hu}$ are not modified by coarse-graining.
Their renormalization flows are then given by 
\begin{equation}
\left\{ 
\begin{array}{l}
\frac{\partial \lambda _{hh}}{\partial l}=(z_{h}+\zeta _{h}-2)\lambda _{hh}%
\nonumber \\ 
\frac{\partial \lambda _{hu}}{\partial l}=(z_{\rho}+\zeta _{h}-2)\lambda
_{hu}%
\end{array}%
, \right.
\end{equation}
where $z_{h}$ and $z_{\rho}$ are the dynamical exponents associated to the
fields $h$ and $\rho$, respectively, and $\zeta _{h}$ is the height
roughness exponent. If a fixed point occurs at $\lambda _{hh}=\lambda
_{hu}\not=0$, the relation 
\begin{equation}
z_{h}+\zeta _{h}=z_{\rho }+\zeta _{h}=2
\end{equation}%
holds exactly, while stability of a fixed point with $\lambda _{hh}=\lambda
_{hu}=0$ requires 
\begin{equation}
z_{h}+\zeta _{h}\leq 2\quad \mathrm{and}\quad z_{\rho }+\zeta _{h}\leq 2.
\end{equation}

\begin{acknowledgments}
This work was supported by the Swiss National Science Foundation through a
Young Researcher Grant and the Harvard University Society of Fellows (RS),
and the NSF through Grant No. DMR-01-18213 (MK).
\end{acknowledgments}

\appendix

\section{Non-Gaussian Steady States}

In this Appendix, we show through a simple example how the method used in
the bulk of the paper may be extended to search for non-Gaussian steady
states. We focus on equations of the form of Eq.~(\ref{cross-gen}) in Sec.~%
\ref{2d}, where the usual Laplacian smoothing term is promoted to a more
general object.

In parallel, we promote $\mathcal{Q}$ (see Eq.~(\ref{q})) to include higher
order terms, as%
\begin{eqnarray}
\mathcal{Q} &=&\int d^{d}x\left( \nu _{ij\alpha \beta }\partial _{\alpha
}h_{i}\partial _{\beta }h_{j}\right.  \nonumber \\
&&+\pi _{ijk\alpha \beta \gamma }\partial _{\alpha }h_{i}\partial _{\beta
}h_{j}\partial _{\gamma }h_{k}  \nonumber \\
&&\left. +\rho _{ijkl\alpha \beta \gamma \delta }\partial _{\alpha
}h_{i}\partial _{\beta }h_{j}\partial _{\gamma }h_{k}\partial _{\delta
}h_{l}+\cdots \right) ,
\end{eqnarray}
while still imposing the relation of Eq.~(\ref{relating}) between $\mathcal{L%
}$ and $\mathcal{Q}$. Following the procedure of Sec.~\ref{general} in the
special case of a single field and in two dimensions (for the sake of
simplicity), we obtain the additional prescriptions 
\begin{equation}
\pi _{xx\alpha }\lambda _{yy}-2\pi _{xy\alpha }\lambda _{xy}+\pi _{yy\alpha
}\lambda _{xx}=0,
\end{equation}%
for any $\alpha $, and 
\begin{equation}
\rho _{xx\alpha \beta }\lambda _{yy}-2\rho _{xy\alpha \beta }\lambda
_{xy}+\rho _{yy\alpha \beta }\lambda _{xx}=0,  \label{cond-quar}
\end{equation}%
for any $\alpha ,\beta $. Here $\pi $ and $\rho $ have been symmetrized, and 
$\alpha ,\beta $ denote $x$ or $y$.

Let us focus on the quartic term with coefficient $\rho $. For rotationally
symmetric tensors%
\begin{equation}
\lambda _{\alpha \beta }=\lambda \delta _{\alpha \beta },
\end{equation}%
and 
\begin{equation}
\rho _{\alpha \beta \gamma \delta }=\rho (\delta _{\alpha \beta }\delta
_{\gamma \delta }+\delta _{\alpha \gamma }\delta _{\beta \delta }+\delta
_{\alpha \delta }\delta _{\beta \gamma }),
\end{equation}%
Eq.~(\ref{cond-quar}) has only the trivial solution $\lambda =0$. By
contrast, an antisymmetric non-linear term with 
\begin{equation}
\lambda _{xx}=-\lambda _{yy}\equiv \lambda ,\quad \text{and}\quad \lambda
_{xy}=0,
\end{equation}%
allows the non-trivial solution 
\begin{equation}
\left\{ 
\begin{array}{l}
\rho _{xxxx}=\rho _{xxyy}=\rho _{yyyy}\equiv \rho _{1} \\ 
\rho _{xxxy}=\rho _{yyyx}\equiv \rho _{2}/2%
\end{array}%
\right. ,
\end{equation}%
with arbitrary $\rho _{1}$ and $\rho _{2}$. Thus, the equation of evolution%
\begin{eqnarray}
\partial _{t}h &=&\nu \nabla ^{2}h+\rho _{1}[(\nabla h)^{2}\nabla ^{2}h+{%
\partial _{x}}h{\partial _{y}}h{\partial _{x}}{\partial _{y}}h]  \nonumber \\
&&+\rho _{2}[{\partial _{x}}h{\partial _{y}}h\nabla ^{2}h+(\nabla h)^{2}{%
\partial _{x}}{\partial _{y}}h]  \nonumber \\
&&+\frac{1}{2}\lambda \lbrack ({\partial _{x}}h)^{2}-({\partial _{y}}%
h)^{2}]+\eta
\end{eqnarray}
obtains the non-Gaussian steady-state distribution%
\begin{eqnarray}
\mathcal{P}_{\text{quartic}} &=&N\exp \left( -\int dxdy\left\{ \frac{\nu }{2}%
(\nabla h)^{2}\right. \right.  \nonumber \\
&&+\frac{\rho _{1}}{12}[({\partial _{x}}h)^{4}+6({\partial _{x}}h)^{2}({%
\partial _{y}}h)^{2}+({\partial _{y}}h)^{4}]  \nonumber \\
&&+\left. \left. \frac{\rho _{2}}{6}[({\partial _{x}}h)^{3}{\partial _{y}}h+(%
{\partial _{y}}h)^{3}{\partial _{x}}h]\right\} \right) .
\end{eqnarray}
We note that this equation of evolution, too, satisfies the
\textquotedblleft hidden\textquotedblright\ symmetry of Eq.~(\ref{symmetry}).


\begin{thebibliography}{99}
\bibitem{diagonalize} For correlated noise, it is often possible to
diagonalize the correlator by a rotation in $h$-space.
\bibitem{temperature} In comparing with the fluctuation--dissipation
condition given in Refs. \cite{Ertas}, one should account for the values of
the temperatures (noise correlators) which were scaled out to unity in Sec. %
\ref{general}.
\end{thebibliography}
\end{document}